\documentstyle[aps,preprint]{revtex}
\begin{document}
\title{Evolution of Superconductivity and Charge Density
Wave Ordering in the Lu$_5$Ir$_4$(Si$_{1-x}$Ge$_x$)$_{10}$ Alloy System}
\author{Yogesh Singh and S. Ramakrishnan}
\address{Tata Institute of Fundamental Research, Mumbai-400005, India}
\maketitle
\begin{abstract}
\noindent 
The compounds Lu$_5$Ir$_4$Si$_{10}$ and Lu$_5$Ir$_4$Ge$_{10}$ crystallize in the tetragonal Sc$_5$Co$_4$Si$_{10}$ type structure. Lu$_5$Ir$_4$Si$_{10}$ is known to superconduct below 3.9~K and it also exhibits a strongly coupled charge density wave (CDW) transition below 83~K. Lu$_5$Ir$_4$Ge$_{10}$ undergoes a transition into the superconducting state below about 2.4~K without any CDW transition at higher temperatures. Recent Si NMR measurements on polycrystalline samples of Lu$_5$Ir$_4$Si$_{10}$ suggest that there is no energy gap at the Si site across the CDW transition. Thus it is of interest to study the evolution of the superconductivity and the CDW transition when we dope at the Si site with small quantities of Ge. Here we present the evolution of T$_C$ and T$_{CDW}$ with concentration x of Ge in the alloy system Lu$_5$Ir$_4$(Si$_{1-x}$Ge$_x$)$_{10}$ (x~=~0.0, 0.005, 0.01, 0.02, 0.05, 0.1, 0.2, 0.4, 1.0) as estimated from dc susceptibility measurements. We find that the CDW is strongly suppressed with increasing x and there is a simultaneous enhancement of the superconducting transition temperature T$_C$ from 3.9~K for the undoped sample to almost 6.6~K for only 10\% concentration of Ge in the alloy.      
\vskip 1truecm 
\noindent     
Ms number ~~~~~~~~~~~~PACS number:~72.10.Fk, 72.15.Qm, 75.20.Hr, 75.30.Mb\\
\end{abstract}
\newpage
\section{Introduction}
\label{sec:INTRO}
\noindent
The compound Lu$_5$Ir$_4$Si$_{10}$ is a member of a class of ternary transition metal silicides which form in the tetragonal Sc$_5$Co$_4$Si$_{10}$ type structure. It has a superconducting transition temperature of 3.9~K and undergoes a phase transition below 83~K which has been shown to be a strongly coupled charge density wave ordering transition \cite{r1,r2,r3}. High pressure studies by Shelton et. al. showed that the superconductivity was enhanced and the CDW was suppressed progressively by the application of pressure \cite{r2}. This implies an intricate interplay between the two phenomena. From heat capacity and susceptibility measurements, almost a 36\% reduction in the density of states at the Fermi level due to the CDW transition was predicted \cite{r2}. However, a recent Si NMR report on powdered polycrystalline samples did not find any energy gap at the Si site associated with the CDW transition \cite{r4}.  
\par
Given that the CDW is suppressed on the application of pressure and that there is no change at the Si site across the CDW transition, one would expect the isostructural compound Lu$_5$Ir$_4$Ge$_{10}$, which has a larger unit cell, to show a CDW transition at an elevated temperature compared to Lu$_5$Ir$_4$Si$_{10}$. However, Lu$_5$Ir$_4$Ge$_{10}$ is found to show only a superconducting transition below 2.4~K without any CDW ordering at high temperatures.
Thus it is of interest to investigate how the CDW and superconductivity evolve when we dope with small quantities of Ge at the Si site in Lu$_5$Ir$_4$Si$_{10}$. Towards this end we have started a detailed investigation of the superconductivity and CDW ordering in the alloy system Lu$_5$Ir$_4$(Si$_{1-x}$Ge$_x$)$_{10}$. Here we present our dc susceptibility measurements to show the dependence of T$_C$ and the T$_{CDW}$ on concentration x of Ge in the alloy.

\section{EXPERIMENTAL DETAILS}
\label{sec:EXPT}
\noindent
Polycrystalline samples of Lu$_5$Ir$_4$(Si$_{1-x}$Ge$_x$)$_{10}$ with x~=~0.0, 0.005, 0.01, 0.02, 0.05, 0.1, 0.2, 0.4, 1.0 were prepared by arc melting together pieces taken in appropriate proportions from master alloys of the parent compounds Lu$_5$Ir$_4$Si$_{10}$ and Lu$_5$Ir$_4$Ge$_{10}$. The samples were annealed in a sealed quartz tube at 950~$^o$C for 8 days. Powder X-ray diffraction measurements confirmed the structure and the absence of any impurity phases.\\
The superconducting transition temperature T$_C$ and the CDW ordering temperature T$_{CDW}$ were determined by measuring the dc susceptibility using a commercial SQUID magnetometer. 

\section{RESULTS AND DISCUSSION}
\label{sec:RES}
\noindent
In Fig.~1 we show the temperature dependence of the susceptibility for the samples with x~=~0.0, 0.005, 0.01, 0.02, 0.05 and 0.1 between 10 and 300~K to concentrate on those values of x where the CDW is seen. The signature of a CDW in the susceptibility is a diamagnetic drop across the transition which comes about due to the reduction of the density of states at the Fermi surface because of the opening up of a gap at the Fermi surface accompanying the CDW ordering.\\
It can immediately be seen that even small doping concentrations affect the CDW strongly. From an onset temperature of 83~K for the undoped sample Lu$_5$Ir$_4$Si$_{10}$, the CDW starts to shift to lower temperatures and also begins to broaden out considerably as we increase the Ge concentration in the alloy. At a concentration of only 10\%, the CDW has been completely wiped out. We have determined T$_{CDW}$ by peaks in the curves of d($\chi$)/dT vs T.\\
In Fig.~2 we show the phase diagram of T$_C$ vs concentration x for all the samples. We have determined the T$_C$ from measurements of $\chi$ vs T (not shown here) between 2 to 8~K in a field of 10~Oe. We find that the T$_C$ increases very rapidly for small amounts of Ge in the alloy when the CDW is also affected the most. At the value of x~=~0.1 for which the CDW has been completely suppressed, the T$_C$ also stop to increase rapidly and sort of saturates at a value of about 6.5~K before decreasing again for higher values of x where the disorder takes over. For comparison we show in the inset of fig.~2, the variation of T$_{CDW}$ with the concentration x.   
    
\section{CONCLUSION}
\label{sec:CON}
We have measured dc susceptibility of the allow system Lu$_5$Ir$_4$(Si$_{1-x}$Ge$_x$)$_{10}$ for x~=~0.0, 0.005, 0.01, 0.02, 0.05, 0.1, 0.2, 0.4, and 1.0 to investigate the evolution of the superconductivity and CDW transitions with increasing concentration of Ge. We find that the CDW transition is strongly suppressed from 83~K for x~=~0 down to 50~K for x = 0.05. There is no signature of the CDW in $\chi$ for higher values of x. There is a simultaneous enhancement of the superconducting transition temperature from 3.9~K for x~=~0.0 to 6.5~K for x~=~0.1.
Our results indicate that there is a strong interplay and competetion between the superconductivity and the CDW ordering in this compound.  Also, we find that the CDW is suppressed even though we are expanding the lattice which suggests that disorder suppresses the CDW more strongly than pressure. In other words, the CDW transition is more sensitive to disorder than to pressure.

\begin{figure}
\caption{The temperature dependence of the Susceptibility for Lu$_5$Ir$_4$(Si$_{1-x}$Ge$_x$)$_{10}$ for x~=~0.0, 0.005, 0.01, 0.02, 0.05 and 0.1. It can clearly be seen that the onset of the CDW shifts to lower temperatures and also broadens out with increasing Ge concentration. 
\label{fres1}}
\end{figure}
\begin{figure}
\caption{The phase diagram of the critical temperature T$_C$ vs concentration x of Ge in the alloy Lu$_5$Ir$_4$(Si$_{1-x}$Ge$_x$)$_{10}$ . The inset shows the variation of T$_{CDW}$ with concentration x. 
\label{fres2}}
\end{figure}
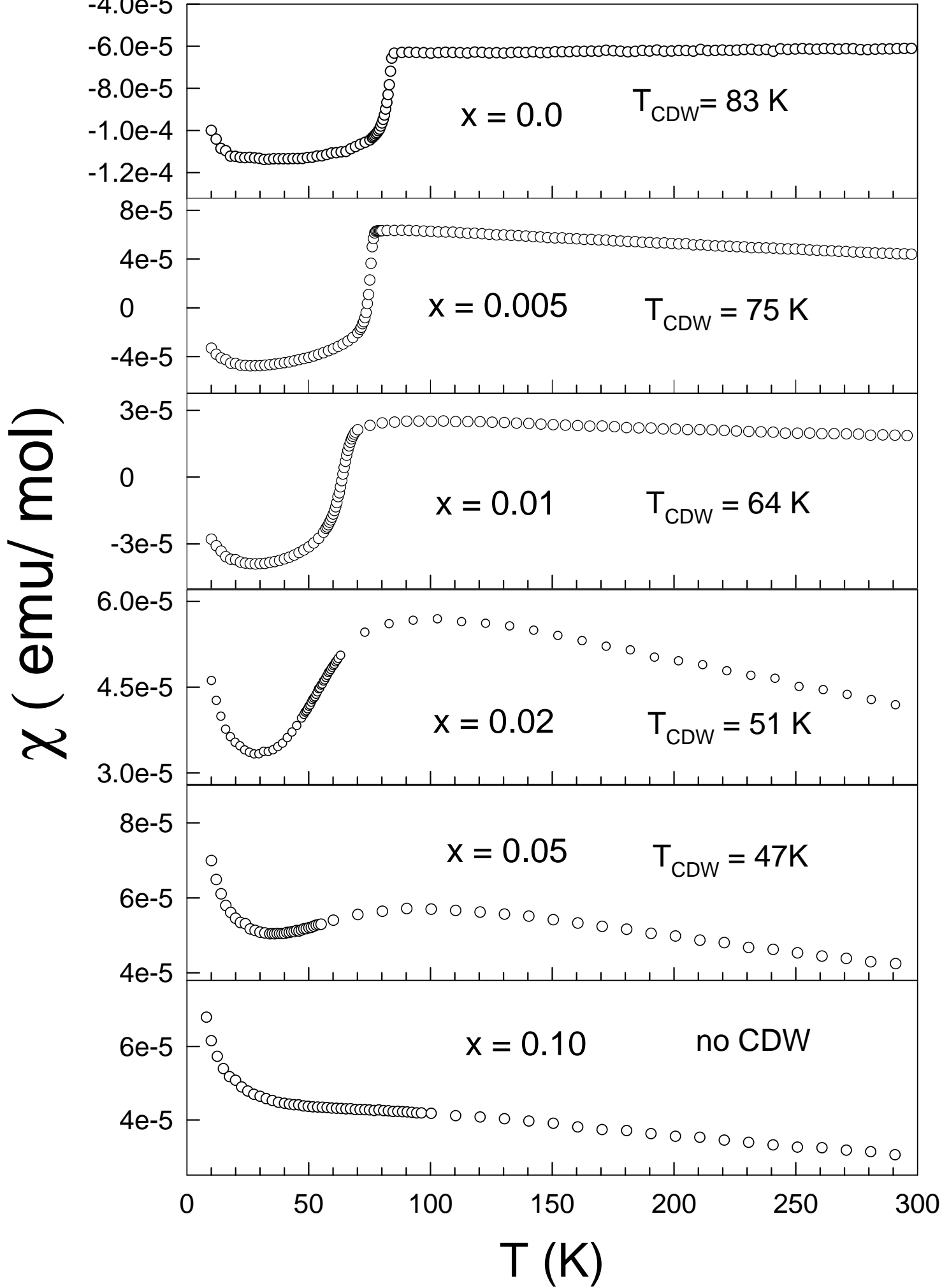
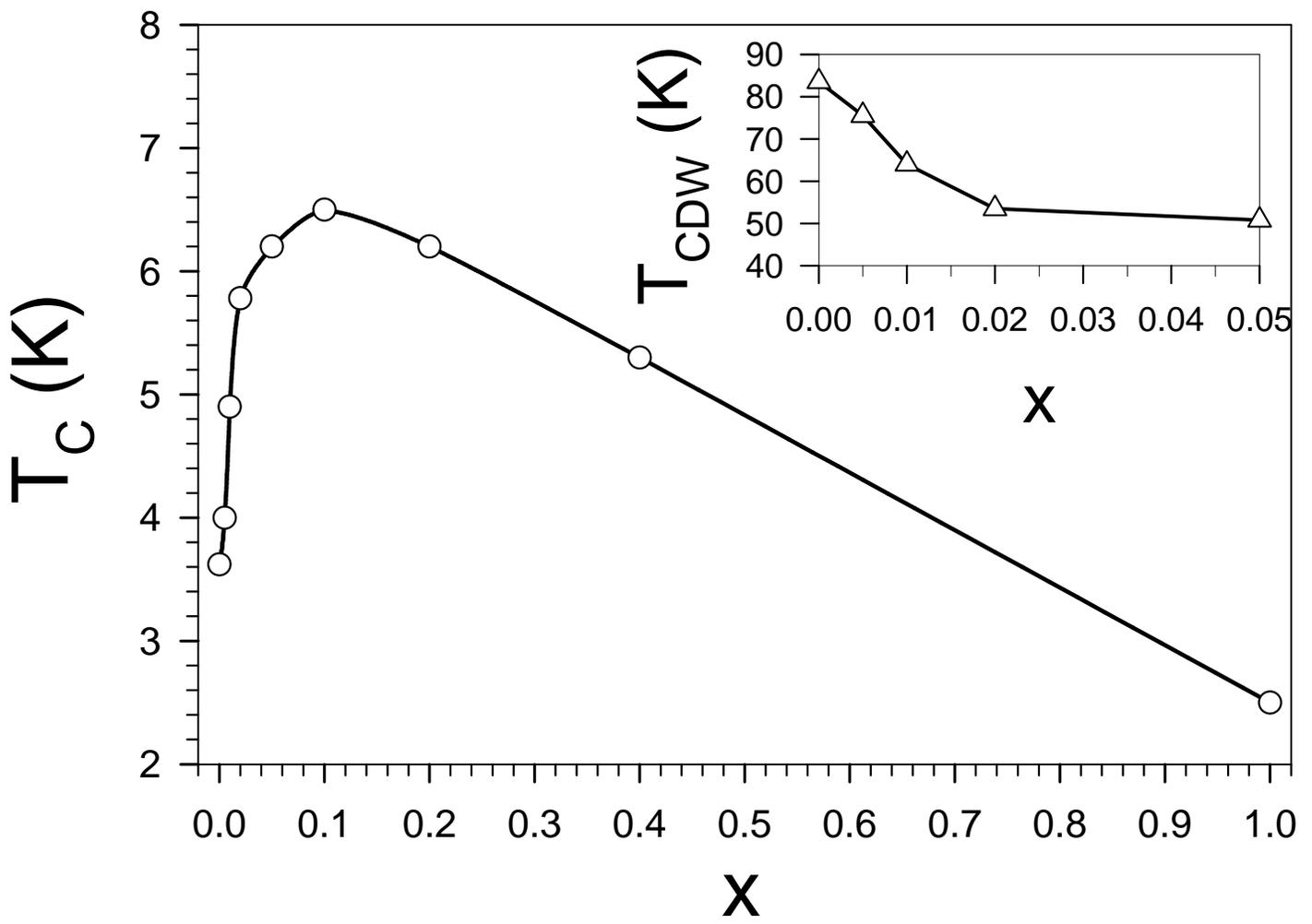
\end{document}